\documentclass[a4paper,11pt]{article}
\usepackage{pos}
\usepackage{hyperref,setspace}

\title{Probing extreme astrophysical accelerators through neutrino anisotropy}
\ShortTitle{Probing extreme astrophysical accelerators through neutrino anisotropy}

\author*[a,b,c]{Marco Stein Muzio}
\author[d,e]{No\'{e}mie Globus}

\affiliation[a]{Department of Astronomy and Astrophysics, Pennsylvania
  State University, University Park, PA 16802, USA} 
\affiliation[b]{Department of Physics, Pennsylvania State University, University Park, PA 16802, USA}
\affiliation[c]{Institute of Gravitation and the Cosmos, Center for Multi-Messenger Astrophysics, Pennsylvania State University, University Park, PA
16802, USA}

\affiliation[d]{Department of Astronomy and Astrophysics, University of California, Santa Cruz, CA 95064, USA}
\affiliation[e]{Astrophysical Big Bang Laboratory, RIKEN, Wako, Saitama, Japan}

\emailAdd{msm6428@psu.edu}

\abstract{We present the extent to which anisotropies in the ultrahigh energy neutrino sky can probe the distribution of extreme astrophysical accelerators in the universe. In this talk, we discuss the origin of an anisotropic neutrino sky and show how observers can use this anisotropy to measure the evolution of ultrahigh energy neutrino sources — and therefore, the sources of ultrahigh energy cosmic rays — for the very first time.}

\ConferenceLogo{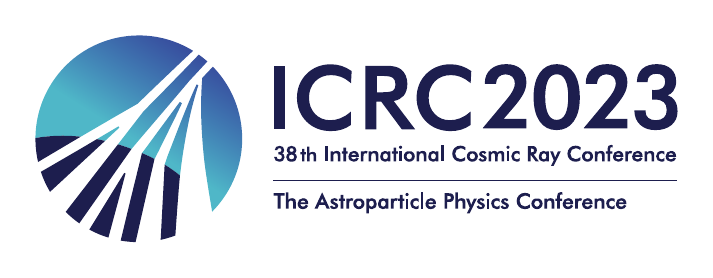}

\FullConference{%
38th International Cosmic Ray Conference (ICRC2023)\\
  26 July - 3 August, 2023\\
  Nagoya, Japan}

\begin{document}
\maketitle

\section{Introduction}

\par
The origins of ultrahigh energy cosmic rays (UHECRs, $E\gtrsim 10^{18}~\mathrm{eV}=1~\mathrm{EeV}$) remain an open question in the field of astroparticle physics. A key piece of evidence towards discovering their origin would be measurement of the distribution of their sources in the universe, i.e.\ their \textit{source evolution}. However, UHECRs observed on Earth probe only a small portion of the universe ($\lesssim 100$~Mpc) due to horizons imprinted by the GZK effect~\cite{Greisen:1966jv,Zatsepin:1966jv} and extragalactic magnetic fields~\cite{Globus:2018svy}. 

\par
Recent investigations into the origin of the dipole in UHECR arrival directions~\cite{PierreAuger:2017pzq} have shown that this dipole can be well-explained if UHECR sources follow the distribution of matter in the universe~\cite{Globus:2018svy,Ding:2021emg}. The local matter distribution (hereafter, the local large scale structure (LSS)) is anisotropic, unlike the homogeneous, isotropic universe at high redshifts. One can understand the UHECR dipole as being an imprint of UHECRs originating only from our local, anisotropic supercluster which is then seen through the lens of magnetic deflections in the Galaxy~\cite{Globus:2018svy}. 

\par
By contrast, neutrinos have no inherent horizon due to their lack of interactions in propagation. In particular, ultrahigh energy neutrinos ($E\gtrsim 30$~PeV) are expected to be produced exclusively through UHECR interactions --- unless one invokes beyond the Standard Model mechanisms~\cite{Ackermann:2022rqc}. Neutrinos can be produced through UHECR interactions both with ambient photons and gas inside their sources (see e.g.~\cite{Muzio:2021zud}), as well as, with the cosmic microwave background (CMB) and extragalactic background light (EBL)~\cite{Greisen:1966jv,Zatsepin:1966jv}. This has two important consequences: first, UHE neutrinos should trace the distribution of UHECR sources, if UHE neutrinos are dominantly produced by in-source UHECR interactions (i.e.\ they are dominantly astrophysical neutrinos); and second, UHE neutrinos will inherit a spectral cutoff from UHECRs, so that UHE neutrinos have an effective maximum energy, $E_\mathrm{max}$. The existence of a maximum energy implies that neutrino sky maps will have an observer-induced horizon imprinted by the energy threshold, $E_\mathrm{th}$, for the map. This horizon is due to neutrinos with energy $E_\mathrm{max}$ redshifting to energies below $E_\mathrm{th}$ if they originate from a redshift $z > z_\mathrm{max} = E_\mathrm{max}/E_\mathrm{th}-1$. This horizon will be more local for higher threshold sky maps, and will therefore probe less of the isotropic high redshift universe and relatively more of the anisotropic local LSS. 

\par
In this study, we aim to explore the degree to which the UHE neutrino sky is anisotropic and how this anisotropy can be used to probe the distribution of UHE neutrino sources (and therefore of UHECRs). Here we layout a framework for such an analysis which could be used by observers and discuss challenges with this approach. 

\section{Model}

\subsection{The density field}\label{sec:densityField}

\par
To begin we make a few simplifying assumptions to calculate the anisotropy of the UHE neutrino sky. First, we assume all sources emit a common neutrino spectral emission rate, $Q_\nu(E)$, and that this astrophysical spectrum is the dominant contribution to the observed UHE neutrino spectrum. Second, we assume that neutrinos from low redshifts follow the local LSS whereas those from the high redshift universe are distributed nearly isotropically. 

\par
The local LSS is, in our calculations, given by the quasi-linear density field based on an ensemble of 20 constrained realizations of the local (within $360$~Mpc) universe~\cite{Hoffman:2018ksb}. A view of this median density field is shown at \href{https://skfb.ly/6AFxT}{https://skfb.ly/6AFxT}. The density field beyond the box boundaries is obtained in the linear regime using a series of linear constrained realisations (based on the linear WF/CRs algorithm~\cite{HoffmanRibak91,Zaroubi:1998di}) within a $1830$~Mpc depth. The use of the linear realisations is justified as the contributions to the anisotropy from beyond the box of $360$~Mpc are dominated by large (linear) scales (see~\cite{Hoffman:2018ksb} for more details). Beyond $1830$~Mpc we assume the density field fluctuations are small enough to be well-approximated as perfectly isotropic.

\subsection{Building skymaps}\label{sec:buildingSkymaps}

Given our assumption that all sources produce a common neutrino spectrum, the integrated flux of neutrinos with energies $E>E_\mathrm{th}$ at Earth which originated from $z<z_0$ is given by 

\begin{align}\label{eq:intFlux}
    J_\nu(E>E_\mathrm{th}, z<z') = \frac{c}{4\pi} \int_0^{z'} \frac{dz}{H(z)} \int_{E_\mathrm{th}}^\infty \mathcal{L}_\nu(z, (1+z)E) dE~,
\end{align}

\noindent
where $\mathcal{L}_\nu(z, E) = \mathcal{H}(z) Q_\nu(E)$ is the neutrino spectral emission rate density and $\mathcal{H}(z)$ is the source density. Defining the local LSS as the structure within the scale at which the universe becomes approximately homogeneous and isotropic, $z_\mathrm{H\&I}$, we can calculate the fraction of the total flux coming from this region $w(z<z_\mathrm{H\&I})$ as 

\begin{align}\label{eq:relFlux}
    w(z<z') = \frac{J_\nu(E>E_\mathrm{th}, z<z')}{J_\nu(E>E_\mathrm{th})}~,
\end{align}

\noindent
where we simplify our notation to $J_\nu(E>E_\mathrm{th})$ for the integrated flux originating from all redshifts. In particular, \eqref{eq:relFlux} depends only on the ratio $E_\mathrm{max}/E_\mathrm{th}$ rather than the absolute energy scale of the cutoff energy. 

\begin{figure}[htpb!]
    \centering
    \includegraphics[width=\textwidth]{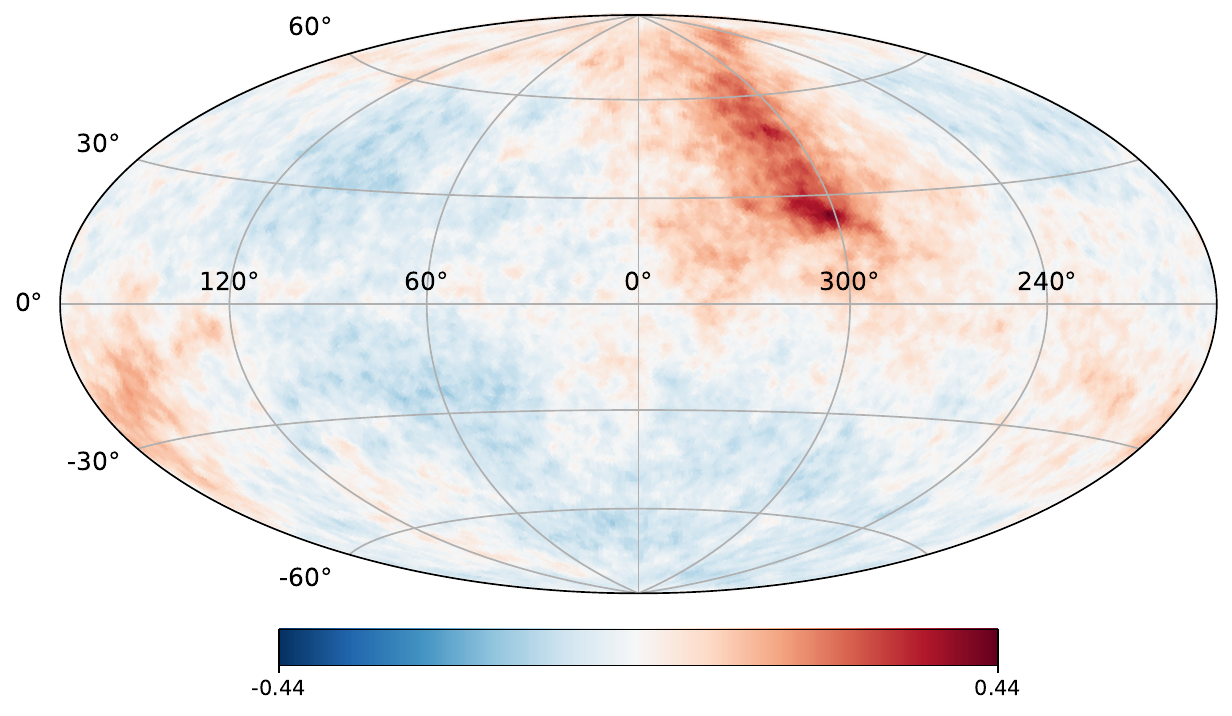}
    \includegraphics[width=\textwidth]{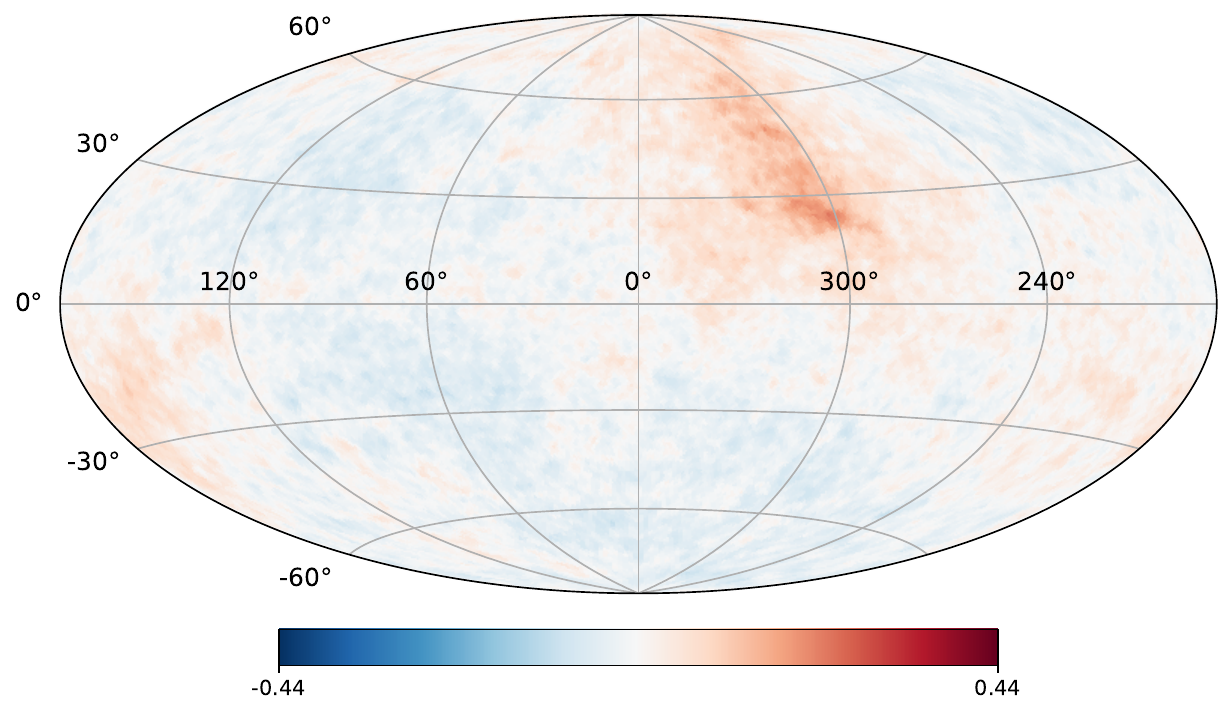}
    \caption{Predicted neutrino sky maps for threshold energies at the maximum neutrino energy for two different source evolution indices: $m=-5$ (top) and $m=0$ (bottom). The color scale indicates the predicted neutrino flux relative to the all-sky average. We assume $z_0=2$ and a spectral index of $\gamma=-2.53$. The top sky map serves as the template map for our maximum likelihood analysis.}
    \label{fig:skymaps}
\end{figure}

\par
Therefore, in order to calculate \eqref{eq:relFlux} we need to know $Q_\nu(E)$ and $\mathcal{H}(z)$. Since neutrinos propagate without significant interactions the neutrino spectrum at Earth is roughly $Q_\nu(E)$ and, therefore, future neutrino observatories will be able to measure it directly. For our purposes we assume that $Q_\nu(E)$ is well-characterized by a single power-law with an exponential cutoff, $Q_\nu(E) \propto E^\gamma \exp(-E/E_\mathrm{max})$. On the other hand, we must adopt a model for the source density, $\mathcal{H}(z)$. For simplicity, we use a two-parameter model of the source evolution

\begin{align}\label{eq:evo}
    \mathcal{H}_{m,z_0}(z) &= \mathcal{H}_0
    \begin{cases} 
      (1+z)^m & z \leq z_0 \\
      (1+z_0)^m e^{-(z-z_0)} & z > z_0
   \end{cases}~.
\end{align}

\par
In order to obtain an accurate estimate of the anisotropy we use the LSS matter density field~\cite{Hoffman:2018ksb} derived CosmicFlows-2 catalogue of peculiar velocities~\cite{Tully:2014gfa}, described in Section~\ref{sec:densityField}. With this field density we can calculate the fraction of the all-sky flux in sky map pixel $i$ as

\begin{align}\label{eq:pixFrac}
    f_i = \sum_j w_j \frac{M_{ij}}{M_j}~,
\end{align}

\noindent
where $w_j = w(z<z_j+\Delta z) - w(z<z_j)$ is the fraction of the total neutrino flux originating from radial shell $j$, $M_j = \rho_j V_j$ is the total mass in a radial shell $j$, and $M_{ij} = \rho_{ij}V_{ij}$ is the total mass in a radial shell $j$ within the solid angle subtended by pixel $i$. Beyond the bounds of the CosmicFlows-2 matter density field we assume $\rho_{ij} = \rho_j$. With this formulation the relative flux from a given direction is determined by the relative amount of matter in that direction, $M_{ij}/M_j$, while the total flux from a given radial shell is determined by the flux fraction, $w_i$. Such a factorization of the radial and angular components is well-justified since even if the number of sources scales the matter density, their inherent luminosity provides an additional scaling of the overall flux from a given radial shell. The product of the inherent source luminosity and number density of sources is fully captured by the source evolution. By contrast, we do not expect the inherent source luminosity to have an angular dependence.

\par
Using \eqref{eq:pixFrac}, the flux in pixel $i$ is simply $J_{\nu,i} = f_i J_\nu$. Example sky maps are shown in Fig.~\ref{fig:skymaps} for two different source evolutions, where we assume $\gamma = -2.53$ (motivated by the best-fit spectral index of the IceCube Cascades dataset~\cite{IceCube:2020acn}), a threshold energy of $E_\mathrm{max}$, and $z_0=2$.

\subsection{Anisotropy measure}

\par
In order to measure the very particular anisotropy arising from the local LSS, we adopt a maximum likelihood analysis. We consider two nested hypotheses: a null hypothesis of an isotropic sky map and an alternative hypothesis of a superposition of an isotropic sky map and a local LSS template. The local LSS template we adopt is the sky map predicted for a very local neutrino horizon, that shown in the top of Fig.~\ref{fig:skymaps} for which $z_\mathrm{max} \simeq 0.11$. Under this framework the number of neutrino events predicted in pixel $i$, $\mu_i$, is given by

\begin{align}
    \mu_i  = \left[\alpha f_i + (1-\alpha) \frac{\Delta \Omega_i}{4\pi}\right] N_\mathrm{evts}~,
\end{align}

\noindent
where $\alpha$ is the relative weight between the two templates ($\alpha=0$ corresponding to the null hypothesis), $\Delta\Omega_i$ is the solid angle subtended by pixel $i$, and $N_\mathrm{evts}$ is the total number of neutrino events in the sky map. Given a set of observed neutrinos $n_i$, the likelihood is then given by the product of Poisson probabilities

\begin{align}
    \mathcal{L}(\alpha| n_i) = \prod_i \mathrm{Poiss}(n_i|\mu_i)~.
\end{align}

\noindent 
With this definition we obtain an estimate of $\alpha$ by maximizing the log-likelihood ratio $\ln\left(\mathcal{L}(\alpha|n_i)/\mathcal{L}(0|n_i)\right)$. In the limit of large statistics, we use this estimate of $\alpha$ to measure the level of neutrino anisotropy for a predicted sky map.

\section{Results}

\par
\begin{figure}
    \centering
    \includegraphics[width=\textwidth]{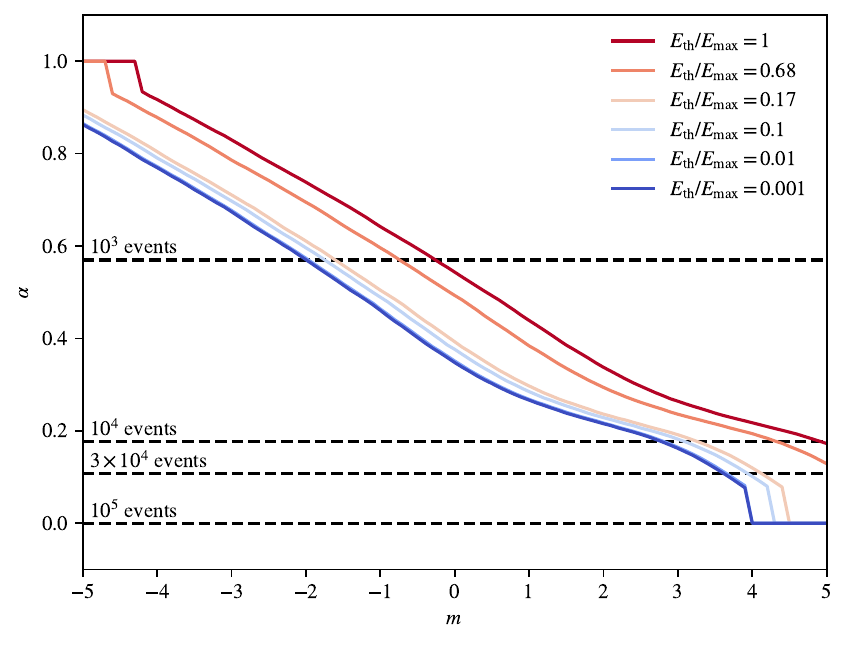}
    \caption{Anisotropy measure $\alpha$ as a function of source evolution power-law index $m$ ($z_0=2$ and $\gamma=-2.53$ fixed) for various threshold energies. Dashed lines indicate the $90\%$ confidence level upper-limit of $\alpha$ given a number of observed events for a truly isotropic distribution.}
    \label{fig:alpha_evo}
\end{figure}

Figure~\ref{fig:alpha_evo} shows the maximum likelihood estimate of $\alpha$ as a function of the source evolution power-law index $m$. As can be seen the value of $\alpha$ is sensitive to the underlying value of $m$. Since the underlying spectrum of neutrinos emitted by the source (i.e. the value of $\gamma$ and $E_\mathrm{max}$) can be directly inferred from observation, measurement of $\alpha$ allows for a measurement of or strong constraint on $m$. In particular, we note Fig.~\ref{fig:alpha_evo} shows that the level of anisotropy increases for increasing energy threshold (i.e. shrinking neutrino horizon), as predicted. This fact allows observers to enhance the level of anisotropy. In fact, the change in anisotropy level with energy threshold can be used to extract additional information about the source evolution, in particular the value of $z_0$, as will be discussed in a forthcoming publication. 

\par
Even if the value of $E_\mathrm{max}$ is unknown, a measurement of (or upper-limit on) $\alpha$ provides a lower-bound on $m$. This is because there is a guaranteed minimum level of anisotropy due to the inherent brightness of the foreground local LSS even for a large neutrino horizon (i.e. small $E_\mathrm{th}/E_\mathrm{max}$ ratio).

\par
The immediate challenge with measuring $\alpha$ comes from the fact that the strength of this anisotropy increases as the threshold energy increases (since this shrinks the observer-induced horizon). However, this increase in energy threshold necessarily decreases the number of neutrino events contributing to the sky map, therefore decreasing the statistical significance of the measurement. Specific analyses must optimize their threshold energy in order to maximize the statistical significance of their measurement. 

\par
\begin{figure}
    \centering
    \includegraphics[width=0.8\textwidth]{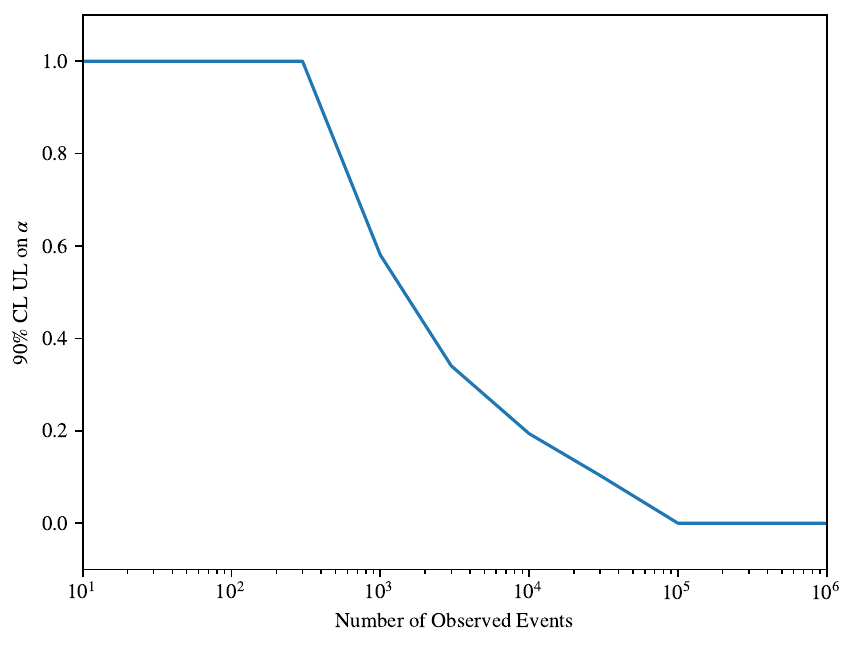}
    \caption{$90\%$ CL upper-limit of the value of $\alpha$ using our maximum likelihood analyses for events sampled from a truly isotropic distribution.}
    \label{fig:UL}
\end{figure}

To give an idea of how many neutrino events are required to measure $\alpha$ Fig.~\ref{fig:UL} shows the $90\%$ confidence level (CL) upper-limit on the value of $\alpha$ given a truly isotropic distribution of neutrino events. This shows that to distinguish from the null hypothesis, analyses require a dataset with $\mathcal{O}(10^3)$ neutrinos. With a $\mathcal{O}(10^5)$ neutrino dataset, the null hypothesis can be rejected at $90\%$ CL for most source evolutions we consider here.

\par
Throughout this analysis we have assumed that the neutrino flux is dominated by astrophysically produced neutrinos. Cosmogenic neutrinos are produced in propagation and, therefore, will be biased towards higher redshift sources. We expect that cosmogenic neutrinos will serve to increase the contribution of high-redshift sources to the total neutrino flux and, therefore, reduce the level of neutrino anisotropy. Similarly, strong extragalactic magnetic fields around filaments may increase the residence time of UHECRs in their source clusters~\cite{Kotera:2008ae}. This would serve to produce an additional neutrino flux contribution which traces the matter distribution of the universe, thereby further increasing the neutrino anisotropy. However, both of these effects are strongly model dependent and is beyond the scope of this current study. 

\par
One final detail which we have not directly addressed is the bias between the normal matter and dark matter density fields. This bias can depend on the properties of the galaxy host to the neutrino source, which may lead some source types to produce neutrino anisotropies which are not well-represented by the simple model described above. We leave investigation of this effect for future work.

\section{Summary}

\par
Ultrahigh energy neutrinos are expected to be produced exclusively by ultrahigh energy cosmic rays. Measurement of the (apparent) end of the ultrahigh energy cosmic ray spectrum implies that, in practice, there is also a maximum neutrino energy. This fact alone implies that neutrinos do not probe the entire universe and, instead, are limited by an observer-induced horizon. 

\par
Throughout this work we have shown that this observer-induced horizon can be used to enhance the anisotropic imprint of the local large-scale structure. Measurement of the level of anisotropy can be used to measure, or strongly constrain, the evolution of the underlying neutrino sources. Since we expect these neutrinos to be produced by ultrahigh energy cosmic rays, this provides an indirect measurement of the evolution of ultrahigh energy cosmic ray sources themselves. Though such a measurement will be challenging to make in practice, it provides a concrete road map to probe the most extreme astrophysical accelerators in the universe.

\acknowledgments

We thank Yehuda Hoffman for his permission to use the
density field from~\cite{Hoffman:2018ksb}. The research of M.S.M. is supported by the NSF MPS-Ascend Postdoctoral Award \#2138121. N.G.'s research is supported by the Simons Foundation, the Chancellor Fellowship at UCSC, and the Vera Rubin Presidential Chair.

\begingroup
\setstretch{0.8}
\setlength{\bibsep}{2.5pt}
\bibliographystyle{ICRC}
\bibliography{references}

\providecommand{\href}[2]{#2}\begingroup\raggedright\begin{thebibliography}{10}

\bibitem{Greisen:1966jv}
K.~Greisen \href{http://dx.doi.org/10.1103/PhysRevLett.16.748}{{\em Phys. Rev.
  Lett.} {\bfseries 16} (1966) 748--750}.

\bibitem{Zatsepin:1966jv}
G.~T. Zatsepin and V.~A. Kuzmin {\em JETP Lett.} {\bfseries 4} (1966) 78--80.

\bibitem{Globus:2018svy}
N.~Globus, T.~Piran, Y.~Hoffman, E.~Carlesi, and D.~Pomar\`ede
  \href{http://dx.doi.org/10.1093/mnras/stz164}{{\em Mon. Not. Roy. Astron.
  Soc.} {\bfseries 484} no.~3, (2019) 4167--4173}.

\bibitem{PierreAuger:2017pzq}
{\bfseries Pierre Auger} Collaboration, A.~Aab {\em et~al.}
  \href{http://dx.doi.org/10.1126/science.aan4338}{{\em Science} {\bfseries
  357} no.~6537, (2017) 1266--1270}.

\bibitem{Ding:2021emg}
C.~Ding, N.~Globus, and G.~R. Farrar
  \href{http://dx.doi.org/10.3847/2041-8213/abf11e}{{\em Astrophys. J. Lett.}
  {\bfseries 913} no.~1, (2021) L13}.

\bibitem{Ackermann:2022rqc}
M.~Ackermann {\em et~al.}
  \href{http://dx.doi.org/10.1016/j.jheap.2022.08.001}{{\em JHEAp} {\bfseries
  36} (2022) 55--110}.

\bibitem{Muzio:2021zud}
M.~S. Muzio, G.~R. Farrar, and M.~Unger
  \href{http://dx.doi.org/10.1103/PhysRevD.105.023022}{{\em Phys. Rev. D}
  {\bfseries 105} no.~2, (2022) 023022}.

\bibitem{Hoffman:2018ksb}
Y.~Hoffman, E.~Carlesi, D.~Pomarede, R.~B. Tully, H.~M. Courtois, S.~Gottlober,
  N.~I. Libeskind, J.~G. Sorce, and G.~Yepes
  \href{http://dx.doi.org/10.1038/s41550-018-0502-4}{{\em Nature Astron.}
  {\bfseries 2} no.~8, (2018) 680--687}.

\bibitem{HoffmanRibak91}
Y.~Hoffman and E.~Ribak \href{http://dx.doi.org/10.1086/186160}{{\em The
  Astrophysical Journal} {\bfseries 380} (09, 1991) L5--L8}.

\bibitem{Zaroubi:1998di}
S.~Zaroubi, Y.~Hoffman, and A.~Dekel
  \href{http://dx.doi.org/10.1086/307473}{{\em Astrophys. J.} {\bfseries 520}
  (1999) 413--425}.

\bibitem{Tully:2014gfa}
R.~B. Tully, H.~Courtois, Y.~Hoffman, and D.~Pomar\`ede
  \href{http://dx.doi.org/10.1038/nature13674}{{\em Nature} {\bfseries 513}
  no.~7516, (2014) 71}.

\bibitem{IceCube:2020acn}
{\bfseries IceCube} Collaboration, M.~G. Aartsen {\em et~al.}
  \href{http://dx.doi.org/10.1103/PhysRevLett.125.121104}{{\em Phys. Rev.
  Lett.} {\bfseries 125} no.~12, (2020) 121104}.

\bibitem{Kotera:2008ae}
K.~Kotera and M.~Lemoine
  \href{http://dx.doi.org/10.1103/PhysRevD.77.123003}{{\em Phys. Rev. D}
  {\bfseries 77} (2008) 123003}.

\end{thebibliography}\endgroup
\endgroup

\end{document}